\documentclass[12pt]{article}
\usepackage{calrsfs}
\usepackage{calligra}
\usepackage{ucs}
\usepackage[utf8x]{inputenc}
\usepackage{braket}
\usepackage[colorlinks=true, citecolor = blue, linkcolor = ForestGreen]{hyperref}
\usepackage[dvipsnames]{xcolor}
\usepackage{amsmath,amssymb}
\usepackage{graphics,graphicx}
\def\a{\alpha}
\def\b{\beta}

\def\d{\delta}

\def\k{\kappa}

\def\be{\begin{equation}}
\def\ee{\end{equation}}
\def\arr{\begin{array}{rll}}
\def\ea{\end{array}}
\def\bea{\begin{eqnarray}}
\def\eea{\end{eqnarray}}

\def\N2{$N{=}2$}

\def\>{\rangle}
\def\<{\langle}
\def\+{\dagger}
\def\={\ =\ }

\begin{document}
\vspace{0.5cm}
\renewcommand{\thefootnote}{\fnsymbol{footnote}}
\renewcommand{\thefootnote}{\fnsymbol{footnote}}
\begin{titlepage}
\setcounter{page}{0}
\vskip 1cm
\begin{center}
{\LARGE\bf $SU(1,1|N)$ superconformal mechanics  }\\
\vskip 0.5cm
{\LARGE\bf  with fermionic gauge symmetry }\\
\vskip 1cm
$
\textrm{\Large Dmitry Chernyavsky \ }
$
\vskip 0.7cm
{\it
School of Physics, Tomsk Polytechnic University,
634050 Tomsk, Lenin Ave. 30, Russia} \\
{E-mail: chernyavsky@tpu.ru}

\end{center}
\vskip 1cm
\begin{abstract} \noindent
We study superpaticle models with fermionic gauge symmetry on the coset spaces of the $SU(1,1|N)$ supergroup. We first construct $SU(1,1|N)$ supersymmetric extension of a particle on $AdS_2$ possessing the $\kappa$--symmetry. Including angular degrees of freedom and extending this model to a superparticle on the $AdS_2\times \mathbb{CP}^{N-1}$ background with two--form flux, one breaks the $\k$--symmetry down to a fermionic gauge symmetry with one parameter. A link of the background field configuration to the near horizon black hole geometries is discussed.
\end{abstract}

\vskip 1,5cm


\end{titlepage}

\renewcommand{\thefootnote}{\arabic{footnote}}
\setcounter{footnote}0

\noindent

\section{Introduction}

There are several reasons for an abiding interest in $d=1$ superconformal models. On the one hand, they provide a convenient framework for getting insight into the structure of higher dimensional superconformal field theories. On the other hand, such systems arise naturally when studying particle dynamics on near horizon black hole backgrounds \cite{Kallosh} which also links to the $AdS_2/CFT_1$--correspondence \cite{AdS/CFT_1, AdS/CFT_1_1}. In particular, it was argued in \cite{Kallosh, GibbonsTownsend} that superconformal mechanics may provide a microscopic quantum description of extreme black holes.
Motivated by this proposal a plenty of $SU(1,1|2)$  superconformal one--dimensional systems and their $D(2,1;\a)$ extensions  have been constructed \cite{NonlinearRealization_1}-\cite{KKLNS1}.  A related line of research concerns the study of superconformal particles propagating on near horizon black hole backgrounds \cite{Zhou}-\cite{Ch}.

There are several competing approaches to the construction of superconformal mechanics: the superfield approach \cite{Superfield_7}, \cite{Superfield_6}, \cite{Superfield_3}-\cite{Superfield_1}, \cite{Ivanov}, the method of nonlinear realizations \cite{NonlinearRealization_1, Anabalon_Zanelli, Zhou, Ch}, and the canonical formalism (e.g. \cite{Galajinsky_Superparticle, GL, HN})\footnote{See also the construction of superconformal mechanics in the context of string theory via specific reduction from higher dimensional superconformal systems, e.g. \cite{TO}}. Some of the models constructed via different methods can in fact be linked to a super 0-brane, i.e. superparticle possessing $\k$--symmetry \cite{Ch}.
Although the $SU(1,1|2)$ supergroup is central for the proposals in \cite{Kallosh, GibbonsTownsend}, viewed more broadly it is only a particular instance in a chain of the $SU(1,1|N)$ supergroups parametrized by an integer $N$.

The goal of this paper is to construct superparticle models on the coset spaces of the $SU(1,1|N)$ supergroup which hold invariant under the $\kappa$--symmetry or its fraction. Since
the explicit realization relies upon specific properties of spinor representations of the rotation subalgebra in the full superconformal algebra\footnote{For a review of the $\kappa$-symmetry in various contexts see \cite{Sorokin}.}, it seems rather surprising that such symmetry is feasible for generic values of $N$. The analysis is also extended to include angular degrees of freedom which yields a superparticle model on the $AdS_2\times \mathbb{CP}^{N-1}$ background with two--form flux in which case the $\kappa$--symmetry is reduced to a one--parametric fermionic gauge symmetry.

The organization of the paper is as follows. In Sect. 2, we consider the geometrically simplest case of the coset space, whose bosonic part is $AdS_2$. Using the method of nonlinear realizations, we construct an invariant dynamical action. It follows from the requirement of the $\kappa$--symmetry. By imposing the gauge fixing condition, we demonstrate that the model is canonically related to the $SU(1,1|N)$ superparticle models constructed earlier in \cite{Ivanov, GL}. Sect. 3 contains the discussion of generalized Gell--Mann matrices and $su(N)$ algebra. An invariant action with extra angular degrees of freedom is constructed and its reduced $\kappa$--symmetry is analyzed. It is demonstrated that the background field configuration associated with the superparticle satisfies the Einstein--Maxwell equations and is linked to the near horizon black hole geometries. In particular, the instance of $N=2$ reproduces the $\kappa$--symmetric super 0--brane propagating in the near horizon region of the extreme Reissner--Nordstr\"om black hole \cite{Zhou}. Concluding Sect. 4 contains the summary and the outlook. There are three appendices with some technical details.

\section{$SU(1,1|N)$ superparticle on $AdS_2$ background}

\subsection{Invariant action and $\k$--symmetry}
 Consider the supercoset space $G/H$, where $G=SU(1,1|N)$ with the structure relation given in Eq. (\ref{algebra}) of Appendix A, and the stability subgroup $H$ generated by the set of operators $\{D, J_a, M\}$. Bosonic part of this superspace is $AdS_2$. The Lie superalgebra
valued MC one--forms are defined by the conventional relation
\bea\label{MC_1}
&&
\tilde G^{-1}d\tilde G=HL_H+KL_K+D L_D+L_a J_a +M L_M+i\left(L_Q Q+\bar Q L_{\bar Q}+L_S S+\bar S L_{\bar S}\right),
\eea
where $\tilde G$ is an element of $G/H$. Here and in what follows we omit indices belonging to the fundamental representation of $su(N)$ algebra for fermionic one--forms and assume the summation over repeated indices, i.e. $L_Q Q=(L_Q)^j Q_j$.
Given the MC one--form, our goal is to construct an invariant action which enjoys the $\k$--symmetry.

It turns out that the presence of the fermionic generators does not affect the structure of the
transformations which hold the same as in the case of the pure bosonic subalgebra. One can show \cite{Ch} that the only invariant bilinear form is  $L_H L_K$, which will be used below for constructing the kinetic term. Because the MC one--forms on the subgroup $H$ transform as connections (see e.g. \cite{Ortin, Krivonos_lectures}), they can be used for building the Wess--Zumino (WZ) term \cite{Ch}. More precisely, when the subgroup $H$ decomposes into the product of an abelian group and some subgroup, the MC forms on this abelian subgroup transform as abelian connections. Hence, their linear combination fits for constructing an invariant action functional. For the case at hand there are two MC one--forms with this property: $L_D$ and $L_M$.
An invariant action on the coset space thus reads
\be\label{Action_1}
S=-m\int\sqrt{4L_H L_K}-\int(aL_D-bL_M),
\ee
where $m$, $a$ and $b$ are constant parameters.

A conventional way to ensure that a model under consideration possess the $\k$--symmetry relies upon a technically convenient representation for variations of the MC one--forms. In Appendix A we expose such variations for the bosonic MC one--forms. Besides, the $\k$--symmetry requires vanishing of the bosonic variations (see e.g. \cite{MT})
\be
[\d x_H]=[\d x_K]=0.
\ee
Taking into account this condition and using (\ref{MK_Variation}), variation of the action can be brought to the form
\bea\label{Action_Variation}
&&
\d_\k S=2 i\int\left\{m  \sqrt{\frac{L_H} {L_K}} [\d \eta]-[\d \psi]\left(a-ib\frac{N-2}{2N}\right)\right\}L_{\bar S}
\nonumber\\[2pt]
&&
+2i\int \left\{m  \sqrt{\frac{L_K} {L_H}} [\d \psi]-[\d \eta]\left(a+ib\frac{N-2}{2N}\right)\right\}L_{\bar Q}+c.c.,
\eea
where the boundary terms $d[\d x_D]$ and $d[\d x_M]$ have been discarded. Demanding (\ref{Action_Variation}) to vanish, one obtains a system of linear algebraic equations on $[\d\psi]$, $[\d\eta]$ and their conjugates, which yields
\be
 m^2=a^2+\left(\frac{N-2}{2N}\right)^2 b^2.
 \ee

Note that the $\kappa$--symmetry reduces the number of (complex) fermionic degrees of freedom from $2N$ to $N$. To reduce the number of fermionic degrees of freedom in another way one could try to accommodate some part of the fermionic variables in the stability subgroup. However, this spoils transformation properties of the MC one--forms and prevents one from constructing invariant bilinears needed to build the action. Thus, the only way to construct a model with the minimal number of fermionic degrees of freedom is to demand it to be invariant under the $\k$--symmetry transformations. The same reasonings are valid for the model with nontrivial angular degrees of freedom which we discuss in section 3.

 \subsection{Explicit form of the action}
In order to construct the action functional in explicit form, let us  define a coset space element
\be\label{Supercoset_Element_1}
\tilde G=e^{tH}e^{zK}e^{i\left(\psi Q+\bar Q\bar\psi\right)}e^{i\left(\eta S+\bar S\bar\eta\right)}.
\ee
As is known \cite{Sorokin}, the $\k$--symmetry reduces the number of fermionic dynamical degrees of freedom by half. It proves convenient to choose the gauge fixing condition in the form  \footnote{There is a subtlety in consistent choosing the gauge fixing condition for the $\kappa$--symmetry in a way compatible with static solutions \cite{Sorokin_Gauge}. It can be verfied that the gauge fixing condition (\ref{Gauge_Fixing_1}) is an appropriate one and can be used without loss of generality.}
\be\label{Gauge_Fixing_1}
\eta=\bar\eta=0.
\ee
As the next step, we construct the MC one--forms
\bea\label{MC_Forms_1}
&&
L_H=dt-i(\psi d\bar\psi-d\psi\bar\psi)+L_K(\psi\bar\psi)^2,
\nonumber\\[2pt]
&&
L_K=z^2dt+dz, \qquad L_D=2z dt, \qquad L_M=-\frac{N-2}{N}L_K\psi\bar\psi.
\eea
For what follows it proves convenient to redefine the coordinates
\be
t\rightarrow t+\frac{1}{z}, \qquad \psi\rightarrow \frac{\psi}{z}, \qquad \bar\psi\rightarrow \frac{\bar\psi}{z},
\ee
which bring the action (\ref{Action_1}) to the form
\be\label{Action_Explicit}
S=-2m\int\sqrt{z^2-\dot z-i(\psi\dot{\bar\psi}-\dot\psi\bar\psi)+(\psi\bar\psi)^2}-2z a-b\frac{N-2}{N}\psi\bar\psi.
\ee

To better understand the structure of the model, let us consider it in the canonical formalism. The Hamiltonian reads
\be\label{Hamiltonian_1}
H=\frac{m^2}{p_z}+z^2p_z+2za+p_z(\psi\bar\psi)^2+b\frac{N-2}{N}\psi\bar\psi,
\ee
where $p_z$ is the momentum canonically conjugate to the bosonic variable $z$.
Fermionic canonical momenta $p_\psi$ and $p_{\bar\psi}$ defined with the use of the right derivatives lead to the second class constraints
 \be
 p_\psi-ip_z\bar\psi=0,\qquad p_{\bar\psi}-ip_z\psi=0.
 \ee
 In order to put the Hamiltonian into the standard conformal mechanics form, let us implement the canonical transformation \cite{Ch}
\be\label{Psi_Redefinition}
\psi\rightarrow \frac{\psi}{\sqrt{2p_z}}, \qquad p_{\psi}\rightarrow \sqrt{2p_z}p_{\psi},
\ee
along with
 \be\label{CanonicalTransformation}
z\rightarrow -\frac{p}{x}-\frac{2 a}{x^2}, \qquad p_z\rightarrow \frac{x^2}{2}.
\ee
Then the Hamiltonian and the constraints take the form
\bea\label{Hamiltonian and Constraints}
&&
H=\frac{p^2}{2}+\frac{b^2}{x^2}\left(\frac{N-2}{N}\right)^2+\frac{2b}{x^2}\frac{N-2}{N}\psi\bar\psi+\frac{2}{x^2}(\psi\bar\psi)^2,
 \nonumber\\[2pt]
&&
p_\psi-\frac{i}{2}\bar\psi=0, \qquad p_{\bar\psi}-\frac{i}{2}\psi=0.
\eea
This is an $SU(1,1|N)$ supersymmetric extension of the standard conformal mechanics.
It should be noted that the original action (\ref{Action_Explicit}) involved two independent parameters, while, as a result of the canonical redefinition, the final Hamiltonian depends only on one. One can verify that the Hamiltonian and constraints (\ref{Hamiltonian and Constraints}) reproduce the one--particle model of the $SU(1,1|N)$ superconformal mechanics in \cite{GL}. Turning to the Lagrangian formalism, our model links to the superparticle constructed within the superfield formalism in pioneering work \cite{Ivanov}. Note that the symmetry structure prompts one to suggest that there should be a relation between these models and a superparticle in the Lobachevsky space \cite{HN}.
\vspace{0.5cm}

\section{Incorporating angular degrees of freedom}

\subsection{Generalized Gell--Mann matrices and $su(N)$ algebra}

In order to generalize our superparticle model by extending it with angular degrees of freedom, it proves convenient to use the fundamental representation matrices given in the bra--ket notations (see e.g. \cite{Gell-Mann_2}). Let us split the set of $(N^2-1)$ traceless hermitian matrices $\lambda_a$ in three subsets $\{T^+_{jk},T^-_{jk},\Lambda_l\}$ such that

$\bullet$ $N(N-1)/2$ symmetric matrices
\be\label{su(N)_S}
T^+_{jk}=\ket{j}\bra{k}+\ket{k}\bra{j}, \qquad j,k=1,\dots, N, \quad j\neq k,
\ee

$\bullet$ $N(N-1)/2$ antisymmetric matrices
\be\label{su(N)_A}
T^-_{jk}=-i\ket{j}\bra{k}+i\ket{k}\bra{j}, \qquad j,k=1,\dots, N,
\ee

$\bullet$ $(N-1)$ traceless diagonal matrices
\be\label{su(N)_D}
\Lambda_l=\sqrt{\frac{2}{l(l+1)}}\left(\sum_{j=1}^l\ket{j}\bra{j}-l\ket{l+1}\bra{l+1}\right), \qquad l=1,\dots,N-1.
\ee
Using the bra--ket notations it is easy to establish the structure relations of $su(N)$. The set of antisymmetric matrices $T^-$ defines the $so(N)$ subalgebra
\be\label{Commutator-}
[T^-_{jk}, T^-_{pq}]=i\left(T^-_{jp}\d_{kq}-T^-_{jq}\d_{kp}-T^-_{kp}\d_{jq}+T^-_{kq}\d_{jp}\right).
\ee
The commutator of $T^+$ yields $T^-$
\be\label{Commutator+}
[T^+_{jk}, T^+_{pq}]=i\left(T^-_{jq}\d_{kp}+T^-_{jp}\d_{kq}+T^-_{kq}\d_{jp}+T^-_{kp}\d_{jq}\right),
\ee
while the mixed commutators read
\be\label{Commutator+-}
[T^+_{jk}, T^-_{pq}]=i\left(T^+_{jp}\d_{kq}-T^+_{jq}\d_{kp}-T^+_{kq}\d_{jp}+T^+_{kp}\d_{jq}\right)+2i (\d_{kq}\d_{jp}-\d_{jq}\d_{kp})\big(\ket{j}\bra{j}-\ket{k}\bra{k}\big).
\ee
Using the fact that (\ref{su(N)_D}) along with the unity matrix define a basis in the space of diagonal $N\times N$ matrices, one can establish the identity \cite{Gell-Mann_2}
\be
\ket{j}\bra{j}=\frac{1}{N}-\sqrt{\frac{j-1}{2j}} \Lambda_{j-1}+\sum_{s=0}^{N-j-1}\frac{\Lambda_{j+s}}{\sqrt{2(j+s)(j+s+1)}}.
\ee
This allows one to rewrite the second term in (\ref{Commutator+-}) in terms of the diagonal traceless matrices $\Lambda_l$
\be
\ket{j}\bra{j}-\ket{k}\bra{k}=\sqrt{\frac{k-1}{2k}}\Lambda_{k-1}-\sqrt{\frac{j-1}{2j}}\Lambda_{j-1}+\sum_{s=j}^{k-1}\frac{\Lambda_s}{\sqrt{2s(s+1)}}, \qquad j<k.
\ee
As the nest step, let us compute the commutator of  $\Lambda_l$ and $T^{\pm}_{jk}$
\be\label{Commutator+-Lambda}
\sqrt{\frac{l(l+1)}{2}}[\Lambda_l, T^{\pm}_{jk}]=\pm i\sum_{s=1}^l \left(T^{\mp}_{sk}\d_{sj}\pm T^{\mp}_{sj}\d_{sk}\right)\pm iT^{\mp}_{jk}\left((k-1)\d_{k,l+1}-(j-1)\d_{j,l+1}\right).
\ee
Finally, since $\Lambda_l$ are diagonal matrices, their commutators vanish. To summarize, (\ref{Commutator-})-(\ref{Commutator+-Lambda}) define the structure relations of $su(N)$ \footnote{As follows from (\ref{Commutator_Matrix}), in order to bring the commutation relations (\ref{Commutator-})-(\ref{Commutator+-Lambda}) to the standard form, one has to multiply each generator by $-i/2$.}.

In the basis chosen it is easy to extract the $su(N-1)$ subalgebra. It is readily verified that the set of operators $\{T^+_{m n}, T^-_{m n}, \Lambda_s\}$, with $m,n=1,\dots, N-1$, $s=1,\dots, N-2$ generates $su(N-1)$. For what follows it proves convenient to introduce the notation
\be\label{GeneratorsCoset}
T^{\pm}_{mN} := T^{\pm}_{m}, \qquad m=1,\dots, N-1.
\ee
As demonstrated in Appendix A, the conventional $su(N)$ commutation relations and (\ref{Commutator-})-(\ref{Commutator+-Lambda}) differ by a factor of $2i$ on the right hand side. In what follows we assume that the duals to the generators $T^{\pm}_{ij}$ and $\Lambda_l$, the MC one--forms $L^{\pm}_{ij}$ and $L_l$ obey the $su(N)$ algebra in the standard form.

\subsection{Invariant action}

In this section we construct an invariant action on the coset space $\frac{SU(1,1|N)}{SO(1,1)\times SU(N-1)\times [U(1)]^2}$ thus generalizing (\ref{Action_1}) to include the angular degrees of freedom. As before, we assume that the first factor in the stability subgroup, $SO(1,1)$, is generated by the dilatation operator $D$. In accordance with the results of the previous section, we set the second factor to be generated by the operators $\{T^+_{m n}, T^-_{m n}, \Lambda_s\}$, where $m,n=1,\dots, N-1$, $s=1,\dots, N-2$. One copy of $U(1)$ in the third factor corresponds to the operator $\Lambda_{N-1}$, while another to $M$.
The remaining bosonic operators $H$, $K$, $T^\pm_m$, the fermions $Q$, $S$ and their conjugate partners $\bar Q$, $\bar S$ generate supercoset space.
Such a choice of the coset identifies the bosonic part with $AdS_2\times \mathbb{CP}^{N-1}$, where $\mathbb{CP}^{N-1}$ is the complex projective space $\mathbb{CP}^{N-1}=\frac{SU(N)}{ SU(N-1)\times U(1)}$.
Let us choose the following parametrization of the coset space
\be\label{Coset}
\tilde G=e^{tH}e^{i\left(\psi Q+\bar Q\bar\psi\right)}e^{zK}e^{i\left(\eta S+\bar S\bar\eta\right)}u,
\ee
where $u$ is an element of $\mathbb{CP}^{N-1}$ generated by $T^{\pm}_m$.

In order to construct an invariant action, one can use (\ref{Action_1}) as an ansatz and extend it by angular degrees of freedom. The corresponding kinetic term can be build from the MC one--forms $L^{\pm}_m$ associated with the generators $T^{\pm}_m$ in (\ref{GeneratorsCoset}) (for $N=2$ see \cite{Ch})
\be\label{CP^N Metric}
L^+_m L^+_m+L^-_m L^-_m.
\ee
Recall that above we constructed the WZ--term using the MC one--forms which transform as abelian connections. In the present case, in addition to $L_D$ and $L_M$ one reveals $L_{N-1}$ possessing the same property.

To summarize, the invariant action on the coset space reads
\be\label{Action_2}
S=-m\int \sqrt{4L_H L_K-L^+_mL^+_m-L^-_mL^-_m}-\int \left(aL_D-bL_M+cL_{N-1}\right),
\ee
where $m$, $a$, $b$ are $c$ are constant parameters. This action describes a supersymmetric extension of a particle on the $AdS_2 \times \mathbb{CP}^{N-1}$ background with two--form flux. Supersymmetric extensions of $\mathbb{CP}^N$ mechanics were earlier studied in \cite{NR_3, BKS, BN}.

\subsection{Reduced $\kappa$--symmetry}
The action functional (\ref{Action_2}) generalizes the model (\ref{Action_1}) which possesses the $\kappa$--symmetry.
For $N=2$ it reproduces a super 0--brane model \cite{Zhou, Ch}. Let us discuss the issue of the $\kappa$--symmetry for $N>2$. Varying the action (\ref{Action_2}), setting
\be\label{Variation_Zero}
[\d x_H]=[\d x_K]=[\d \theta^\pm_m]=0,
\ee
and proceeding along the same lines as above,  one obtains a system of the algebraic equations
\bea\label{LinearEquation}
&&
\frac{m\left(2i L_H[\d \eta]+[\d \psi] T^+_{m}L^+_m+[\d \psi] T^-_{m}L^-_m\right)}{\sqrt{4L_H L_K-L^+_mL^+_m-L^-_mL^-_m}}-i[\d\psi] \left(a-i b \frac{N-2}{2N}-i c \Lambda_{N-1}\right)=0,
 \nonumber\\[2pt]
&&
\frac{m\left(2i L_K[\d \psi]-[\d \eta] T^+_{m}L^+_m-[\d \eta] T^-_{m}L^-_m\right)}{\sqrt{4L_H L_K-L^+_mL^+_m-L^-_mL^-_m}}-i[\d\eta] \left(a+i b \frac{N-2}{2N}+i c \Lambda_{N-1}\right)=0,
\eea
as well as for the complex conjugate pair $[\d \bar\psi]$, $[\d \bar\eta]$. In order to see if the system admits a solution,  let us express $[\d\psi]$ from the second equation and substitute it into the first. This gives a linear equation on $[\d\eta]$ which decomposes into four independent linear equations. The first two of them are proportional to the contraction with the MC one--forms $L^{\pm}_m$
\be
b\frac{N-2}{N}[\d\eta]T^{\pm}_m+c[\d\eta]\{\Lambda_{N-1}, T^{\pm}_m\}=0,
\ee
where curly bracket stands for the anticommutator. Taking into account the form of the matrices (\ref{GeneratorsCoset}) in the bra--ket notations (\ref{su(N)_S})-(\ref{su(N)_D}), one can establish the identity
\be
 \{\Lambda_{N-1}, T^{\pm}_m\}=-\sqrt{\frac{2}{N(N-1)}}(N-2)T^{\pm}_m.
\ee
Hence, the constant parameters $b$ and $c$ should be related to each other
\be\label{Restriction_1}
b=c\sqrt{\frac{2N}{N-1}}.
\ee
The next equation is proportional to the bilinear form $L_H L_K$ and reads
\be\label{Equation_3}
[\d\eta]-\frac{[\d\eta]}{m^2}\left(a^2+b^2\left(\frac{N-2}{2N}\right)^2+ c^2\Lambda_{N-1}^2+b c\frac{N-2}{N}\Lambda_{N-1}\right)=0.
\ee
Again, using the bra--ket notations one can find
\be
\Lambda_{N-1}^2=\frac{2}{N(N-1)}\sum_{j=1}^{N-1}\ket{j}\bra{j}+\frac{2(N-1)}{N}\ket{N}\bra{N}.
\ee
In order to satisfy this equation with nontrivial $c$, either the first $(N-1)$ components of $[\d\eta]$ should be vanishing or the last one. Both cases imply additional restrictions on the parameters. Assuming that (\ref{Equation_3}) holds, the last equation coming from (\ref{LinearEquation}) reads
\bea\label{Equation_4}
&&
[\d\eta]\left(L^+_m T^+_m+L^-_m T^-_m\right)\left(L^+_n T^+_n+L^-_n T^-_n\right)-[\d\eta]\left(L^+_mL^+_m+L^-_mL^-_m\right)=0.
\eea
In this expression one encounters the following anticommutators
\bea\label{Anticommutator}
&&
\{T^+_m,T^+_n\}=T^+_{mn}+2\d_{mn}\ket{N}\bra{N}, \qquad\{T^+_m,T^-_n\}=-T^-_{mn}.
 \nonumber\\[2pt]
&&
\{T^-_m,T^-_n\}=T^-_{mn}+2\d_{mn}\ket{N}\bra{N}, \qquad m,n=1,\dots, N-1.
\eea
Note that the matrices $T^\pm_{mn}$ act in the $(N-1)$--dimensional space of vectors. It follows from (\ref{Equation_4}) that its solution cannot have the first $(N-1)$ nontrivial components which leaves one with
\be
\bra{[\d \eta]}=\kappa \bra{N},
\ee
where $\kappa$ is an anticommuting single complex gauge parameter. In view of (\ref{Equation_3}) the additional restriction on the parameters, which was mentioned above, reads
\be\label{Restriction_2}
m^2=a^2+\frac{c^2N}{2(N-1)}.
\ee

To summarize, we conclude that the action (\ref{Action_2}) possesses a reduced $\kappa$--symmetry with a single fermionic gauge parameter provided the restrictions (\ref{Restriction_1}) and (\ref{Restriction_2}) hold. Note that the case of $N=2$ reveals a subtlety. For $N=2$ the matrices $T^{\pm}_{mn}$ in (\ref{Anticommutator}) are vanishing and there exists a solution with two gauge parameters which yields the conventional  $\kappa$--symmetry (see \cite{Zhou, Ch}).

\subsection{Background geometry and bosonic part of the action}
The MC one--forms used in the construction of the action (\ref{Action_2}) are exposed in Appendix B. One can impose a gauge fixing condition by requiring the $N$-th component of the fermionic variables $\eta$ and $\bar\eta$ to be vanishing. The gauge fixed action has a complicated form and in what follows we focus on its bosonic part only.

Interestingly enough, the bosonic part of the action (\ref{Action_2}) can be interpreted as a particle propagating in external gravitational and electromagnetic fields.  In order to understand whether this field configuration satisfies the Einstein--Maxwell equations, let rewrite the metric in $AdS$--basis by making use of the coordinates redefinition
\be
t\rightarrow \frac{1}{2}\left(t+\frac{1}{r}\right), \qquad z\rightarrow r.
\ee
In these coordinates the metric and the gauge field one--form read
\bea\label{Metric}
&&
ds^2=\gamma^2\left(r^2 dt^2-\frac{dr^2}{r^2}-L^+_{m}L^+_{m}-L^-_{m}L^-_{m}\right),
 \nonumber\\[2pt]
&&
A=\a r dt+\b L_{N-1},
\eea
where  $L^{\pm}_i$, $L_{N-1}$ are the MC one--forms (\ref{MKFroms}) with the fermionic parts discarded, while $\a$, $\b$ and $\gamma$ are some constant parameters related to $a$, $b$ and $c$ in (\ref{Action_2}). The metric above describes $2N$--dimensional space $AdS_2\times \mathbb{CP}^{N-1}$.
The Maxwell two--form can be found by using the MC equations (\ref{MCEquations}) and the commutation relations (\ref{Commutator+-})
\be\label{Maxwell_Form}
F=-\a dt\wedge dr-\b \sqrt{\frac{N}{2(N-1)}} L^+_m\wedge L^-_m.
\ee
Let us prove that this two--form satisfies the Maxwell equations
\be
d \ast F=0,
\ee
where $\ast$ is the Hodge dual operator. The dual form to the first term in (\ref{Maxwell_Form}) is proportional to the volume form on $\mathbb{CP}^{N-1}$ and hence it is closed. The Hodge dual of the second term is proportional to the exterior product of $dt\wedge dr$ and the linear combination of $2(N-2)$-forms on $\mathbb{CP}^{N-1}$. Clearly, the first term of this product is closed. In order to prove that the linear combination is closed as well, one has to use the MC equations
\bea
&&
dL^{\pm}_m=L^{\pm}_{mq}\wedge L^-_q\pm L^\mp_{mq}\wedge L^{+}_q\pm\sum_{l=m}^{N-1}\sqrt{\frac{1}{2l(l+1)}}L_l\wedge L^{\mp}_m
 \nonumber\\[2pt]
&&
\qquad \qquad \qquad \qquad \qquad\qquad\qquad \mp\sqrt{\frac{m-1}{2m}}L_{m-1}\wedge L^{\mp}_m \pm \sqrt{\frac{N-1}{2N}}L_{N-1}\wedge L^{\mp}_m.
\eea
From these equations one concludes that (\ref{Maxwell_Form}) satisfies the Maxwell equations without imposing any restrictions on the constant parameters $\a$ and $\b$.

It proves convenient to analyze the Einstein equations in tetrad formalism. In Appendix B the geometrical characteristics of $\mathbb{CP}^{N-1}$ are given. Using those results one can verify that the Einstein equations
\be
R_{ab}-\frac{R}{2}\eta_{ab}-2\left(F_{ac}F_{bc}-\frac{1}{4}F^2\eta_{ab}\right)=0,
\ee
where $\eta_{ab}$ is an $2N$--dimensional Minkowski metric, hold provided the constant parameters obey the restrictions
\bea
&&
\frac{\gamma^2}{2}(N+2)=2\a^2+\frac{\beta^2 N}{N-1},
 \nonumber\\[2pt]
&&
\gamma^2\left(1+(N-2)(N+1)\right)=\a^2+\beta^2\frac{N}{2}.
\eea

Notice that the case of $N=2$ reveals a subtlety. The two equations above reduce to $\gamma^2=\a^2+\beta^2$ and the resulting solution (\ref{Metric}) describes the near horizon region of the extreme Reissner--Nordstr\"{o}m black hole \cite{Zhou,Ch}.
It is natural to wonder whether the geometries associated with the $N>2$ models are linked to black hole configurations as well. Recall that
$(2N-1)$--dimensional sphere can be presented as a Hopf fibration over $\mathbb{CP}^{N-1}$ and the corresponding metric reads (see e.g. Appendix B of \cite{HopfFibration})
\be\label{SphereMetric}
d\Omega^2_{2N-1}=L^+_{m}L^+_{m}+L^-_{m}L^-_{m}+(d\psi+L_{N-1})^2.
\ee
Geometry in the near horizon region of a generic spherically symmetric charged black hole solution is represented by a product of $AdS_2$ and a sphere. On can verify that in an odd--dimensional space this geometry can be reduced in the $\psi$ direction thus giving the configuration of fields (\ref{Metric}). For a particle propagating on such background, the only effect of this reduction is the fixation of the momentum conjugate to the coordinate $\psi$. The above reasoning suggests that the bosonic part of the action (\ref{Action_2}) describes a particle in the near horizon region of a spherically symmetric black hole with the fixed canonical momentum $p_\psi$.

\section{Conclusion}
To summarize, within the framework of the method of nonlinear realizations the $SU(1,1|N)$--invariant particle models have been constructed. Our consideration was
primarily focused on the two different coset spaces of $SU(1,1|N)$ supergroup. First we defined the coset superspace with the bosonic part represented by $AdS_2$ and built a superparticle on it possessing the $\k$--symmetry. Having fixed the gauge, we demonstrated that it is canonically equivalent to the superparticle models of \cite{Ivanov, GL}.
Then we incorporated angular degrees of freedom into the scheme which originated from the $SU(N)$ subgroup. The resulting model describes a supersymmetric extension of a particle on $AdS_2 \times \mathbb{CP}^{N-1}$ space. The particular case of $N=2$ corresponds to the super 0--brane propagating in the near horizon region of the Reissner--Nordstr\"{o}m black hole  \cite{Zhou, Ch}. It was shown that for $N>2$ the $\kappa$--symmetry reduces to a one-parametric fermionic gauge symmetry. This correlates with the analysis in \cite{GL}. The authors of \cite{GL} encountered a problem in constructing an $SU(1,1|N)$ superparticle with angular variables within the canonical formalism. Our analysis suggests that in order to realize $SU(1,1|N)$ symmetry for $N>2$ one has to introduce more fermionic dynamical degrees of freedom. To fully resolve this it seems natural to work within the  superfield formalism.

The bosonic part of the action with angular degrees of freedom was shown to be related to the near horizon black hole geometries with the spherical symmetry. As it is known, the symmetry group of the near horizon Myers-Perry black hole with equal rotating parameters is $SO(1,2)\times SU(N)$. It would be interesting to study a possible link between the superparticles in this work and those geometries.  Finally, it is of interest to generalize our classical treatment and consider the models at the quantum level\footnote{Regarding the quantization of superparticle models on the coset spaces see recent works \cite{HJ, HJ_2} and references therein.}. In particular, it is worth analyzing the role of the $\kappa$-- and reduced $\kappa$--symmetry from the quantum perspective.

\section*{Acknowledgements}
This work was supported by the Tomsk Polytechnic University competitiveness enhancement program, the RF Presidential grant MK-2101.2017.2 and the RFBR grant 18-52-05002.

\appendix
\numberwithin{equation}{section}

\section{$su(1,1|N)$ superalgebra}

In this work we use the notations in \cite{GL} for the structure relations of the superalgebra $su(1,1|N)$
\begin{align}\label{algebra}
&
[ H,D ]=H\ , && [ H,K ]=2D\ ,
\nonumber\\[2pt]
&
[D,K]=K\ , && [ J_a,J_b ]=f_{abc} J_c\ ,
\nonumber\\[2pt]
& [ D,Q_j] = -\frac{1}{2} Q_j\ , && [ D,S_j] =\frac{1}{2} S_j\ ,
\nonumber\\[2pt]
&
[ K,Q_j ] =S_\a\ , && [ H,S_j ]=-Q_j\ ,
\nonumber\\[2pt]
&
[ J_a,Q_j] =\frac{i}{2} {{(\lambda_a)}_j}^k Q_k\ , && [J_a,S_j] =\frac{i}{2} {{(\lambda_a)}_j}^k S_k\ ,
\nonumber\\[2pt]
& [ D,\bar Q^j ] =-\frac{1}{2} \bar Q^j\ , && [ D,\bar S^j] =\frac{1}{2} \bar S^j\ ,
\nonumber\\[2pt]
& [K,\bar Q^j] =\bar S^j\ , && [ H,\bar S^j] =-\bar Q^j\ ,
\nonumber\\[2pt]
&
[J_a,\bar Q^j] =-\frac{i}{2} \bar Q^k {{(\lambda_a)}_k}^j\ , && [ J_a,\bar S^j] =-\frac{i}{2}
\bar S^k {{(\lambda_a)}_k}^j\,
\nonumber\\[2pt]
&
[M,Q_j]=i Q_j,\ && [M,\bar Q^j]=-i\bar Q^j, \
\nonumber\\[2pt]
&
[M,S_j]=i S_j,\ && [M,\bar S^j]=-i\bar S^j, \
\nonumber\\[2pt]
&
\{Q_j,\bar{Q}^k\}=-2iH \d_j{}^k,\ && \{Q_j,\bar{S}^k\}=2(\lambda_a)_j{}^k J_a +\left(2iD-\frac{N-2}{N}M\right)\d_j{}^k,\
\nonumber\\[2pt]
&
\{S_j,\bar{S}^k\}=-2iK \d_j{}^k\,&& \{S_j,\bar{Q}^k\}=-2(\lambda_a)_j{}^k J_a +\left(2iD+\frac{N-2}{N}M\right)\d_j{}^k.\
\end{align}
The bosonic part of the superlagebra is presented by a direct sum of the conformal algebra $so(1,2)$, generated by $H$, $K$, $D$, the $R$--symmetry subalgebra $su(N)\oplus u(1)$, which corresponds to the operators $J_a$ and $M$. Matrices $(\lambda_a)_j{}^k$  define fundamental representation of $su(N)$, i.e. they are  hermitian traceless matrices of the dimension $N\times N$, $j,k=1,\dots, N$ which satisfy the commutation relations
\be\label{Commutator_Matrix}
[\lambda_a,\lambda_b]=2i f_{abc}\lambda_c,
\ee
where, as in (\ref{algebra}), $f_{abc}$ are totally antisymmetric structure constants of $su(N)$. The fermionic complex generators obey the conjugation rules
\be\label{Hermitian_1}
Q_\a^\dag=\bar Q^\a, \qquad S_\a^\dag=\bar S^\a.
\ee

For reader's convenience we display below the MC equations for the bosonic forms
\bea\label{MCEquations}
&&
dL_H=-L_H\wedge L_D-2i L_Q\wedge L_{\bar Q},
\nonumber\\[2pt]
&&
dL_K=L_K\wedge L_D-2i L_S\wedge L_{\bar S},
\nonumber\\[2pt]
&&
dL_D=-2L_H\wedge L_K+2i\left(L_Q\wedge L_{\bar S}+L_S\wedge L_{\bar Q}\right),
\nonumber\\[2pt]
&&
dL_a=-\frac{1}{2}f_{abc}L_b\wedge L_c+2\left(L_Q\lambda_a\wedge L_{\bar S}-L_S\lambda_a\wedge L_{\bar Q}\right),
\nonumber\\[2pt]
&&
dL_M=\frac{N-2}{N}\left(L_S\wedge L_{\bar Q}-L_Q\wedge L_{\bar S}\right).
\eea
Using these equations, variations of the MC one--forms can be put in the form
\bea\label{MK_Variation}
&&
\d L_H=d [\d x_H] + [\d x_D] L_H-L_D[\d x_H]-2i \left([\d \psi]L_{\bar Q}-L_Q[\d \bar\psi]\right),
\nonumber\\[2pt]
&&
\d L_K=d [\d x_K] - [\d x_D] L_K+L_D[\d x_K]-2i \left([\d \eta]L_{\bar S}-L_S[\d \bar\eta]\right),
\nonumber\\[2pt]
&&
\d L_D=d [\d x_D] - 2[\d x_H] L_K+2[\d x_K] L_H+2i  \left([\d \psi]L_{\bar S}-L_Q[\d \bar\eta]+[\d \eta]L_{\bar Q}-L_S[\d \bar\psi]\right), \nonumber
\nonumber\\[2pt]
&&
\d L_a= d[\d x_a]-f_{abc}[\d x_b] L_c+2\left(L_S\lambda_a[\d \bar\psi]-[\d \eta]\lambda_a L_{\bar Q}+[\d \psi]\lambda_a L_{\bar S}-L_Q\lambda_a[\d \bar\eta]\right),
\nonumber\\[2pt]
&&
\d L_M= d[\d x_M]+\frac{N-2}{N}\left([\d \eta]L_{\bar Q}-L_S[\d \bar\psi]-[\d \psi]L_{\bar S}+L_Q[\d \bar\eta]\right),
\eea
where, following \cite{Anabalon_Zanelli}, we introduced the notation
\be
[\d Z^A]=L^A{}_M\d Z^M,
\ee
for a  MC one--form $L^A=L^A{}_M dZ^M$.

\section{Bosonic MC one--forms incorporating angular variables}

The bosonic MC one--forms for the coset element (\ref{Coset}) read
\bea\label{MKFroms}
&&
L_H=Dt,
\nonumber\\[2pt]
&&
L_K=z^2 Dt+dz+Dt(\eta\bar\eta)^2-2\eta\bar\eta\left(d\psi \bar\eta+\eta d\bar\psi\right)-2iz\left(d\psi\bar\eta-\eta d\bar\psi\right)-i(\eta d\bar\eta-d\eta\bar\eta),
\nonumber\\[2pt]
&&
L_D=2zDt+i(\eta d\bar\psi-d\psi\bar\eta),
\nonumber\\[2pt]
&&
L_a=L_a^0+2Dt(\eta\lambda_b\bar\eta)U_{ab}-2\left(d\psi \lambda_b\bar\eta+\eta\lambda_b d\bar\psi\right)U_{ab},
\nonumber\\[2pt]
&&
L_M=\frac{N-2}{N}\left(d\psi\bar\eta+\eta d\bar\psi-\eta\bar\eta Dt\right), \qquad Dt=dt-i(\psi d\bar\psi-d\psi\bar\psi),
\eea
where $L_a^0$ are the bosonic MC one--forms on the coset space $\frac{SU(N)}{SU(N-1)\times U(1)}$
\be
u^{-1}du=L_a^0J_a,
\ee
and the matrix $U_{ab}$ defines a group element of $SU(N)$ in the adjoint representation
\be
u^{-1}J_a u=U_{ab}J_b.
\ee
When obtaining these equations, the identity
\be
\frac{1}{2}(\lambda_a)_{\a}{}^\b(\lambda_a)_\gamma{}^\rho=-\frac{1}{N}\d_{\a}{}^\b \d_\gamma{}^{\rho}+\d_\gamma{}^\beta\d_\a{}^\rho,
\ee
proves to be helpful.
\section{Curvature of $\mathbb{CP}^{N-1}$}
In this Appendix we compute the Ricci tensor and the scalar curvature for the metric on $\mathbb{CP}^{N-1}$. Note that the algebra $su(N)$ can be written in the following form:
\bea
[P_\a,P_\b]=f_{\a\b A}M_A, \qquad [P_\a,M_A]=f_{\a A \b}P_{\b}, \qquad [M_A, M_B]=f_{ABC}M_C,
\eea
where we denoted the stability subgroup generators $T^{\pm}_{mn}$, $\Lambda_l$ for $m,n,l=1\dots, N-1$ collectively by $M_A$, while the remaining operators by $P_\a$. Given the algebra, let us rewrite it in the dual form
\bea\label{Algebra_Dual}
&&
dL_\a+f_{ \a\b A}L_\b\wedge L_A =0,
\nonumber\\[2pt]
&&
dL_A+\frac{1}{2} f_{ABC}L_B\wedge L_C+\frac12 f_{A \a\b}L_{\a}\wedge L_\b=0.
\eea
Let us define the invariant metric on $\mathbb{CP}^{N-1}$ as a quadratic combination (\ref{CP^N Metric}) rewritten in the condensed notations
\be
ds^2=L_\a L_\a.
\ee
Without distinguishing upper and lower indices (which are all Euclidian), one then introduces the tetrad $e^\a=L_\a$ and writes down the equation for the spin connection $\omega^{\a\b}$
\be
d e^\a+\omega^{\a\b}\wedge e^\b=0.
\ee
Using the MC structure relations (\ref{Algebra_Dual}), one finds
\be\label{SpinConnection}
\omega^{\a\b}=-f^{\a\b A}L_A.
\ee
Substituting it into the equation defining the curvature two--form
\be
R^{\a\b}=d\omega^{\a\b}+\omega^{\a\gamma}\wedge \omega^{\gamma\b},
\ee
and using the Jacobi identity for the structure constants, one finds
\be
R^{\a\b}=\frac{1}{2} f^{\a\b A}f^{\gamma \d A}L_\gamma\wedge L_\d.
\ee
Taking into account the explicit form of the $su(N)$ structure constants and using the equation above, one can find the Ricci tensor in the tetrad notation
\be
R_{\stackrel{+}{m}\stackrel{+}{n}}=\frac{N}{2} \d_{\stackrel{+}{m}\stackrel{+}{n}}, \qquad R_{\stackrel{-}{m}\stackrel{-}{n}}=\frac{N}{2} \d_{\stackrel{-}{m}\stackrel{-}{n}},
\ee
while the scalar curvature reads
\be
R=N(N-1).
\ee

\end{document}